\documentclass [11pt]{article}

\usepackage{amsmath}
\usepackage{amssymb}
\usepackage{graphicx}

\newtheorem{exam}{Example}[section]
\newtheorem{lemm}{Lemma}[section]

\newtheorem{theo}{Theorem}[section]

\begin{document}

\title{The Existence of Quantum Entanglement Catalysts\thanks{This work was partly
supported by the National Foundation of Natural Sciences of China
(Grant Nos: 60223004,60496321,60321002, and 60305005).}}
\author{Xiaoming Sun\thanks{Email: sun\_xm97@mails.tsinghua.edu.cn}\ \ \ Runyao Duan\thanks{Email: dry02@mails.tsinghua.edu.cn}\ \ \ Mingsheng Ying\thanks{Email: yingmsh@mail.tsinghua.edu.cn}\\
\\
\small \em
State Key Laboratory of Intelligent Technology and Systems,\\
\small \em Department of Computer Science and Technology, Tsinghua
Univ., Beijing, 100084, China.}

\date{}
\maketitle

\begin{abstract}
Without additional resources, it is often impossible to transform
one entangled quantum state into another with local quantum
operations and classical communication. Jonathan and Plenio [Phys.
Rev. Lett. 83, 3566(1999)] presented an interesting example
showing that the presence of another state, called a catalyst,
enables such a transformation without changing the catalyst. They
also pointed out that in general it is very hard to find an
analytical condition under which a catalyst exists. In this paper
we study the existence of catalysts for two incomparable quantum
states. For the simplest case of $2\times 2$ catalysts for
transformations from one $4\times 4$ state to another, a necessary
and sufficient condition for existence is found. For the general
case, we give an efficient polynomial time algorithm to decide
whether a $k\times k$ catalyst exists for two $n\times n$
incomparable states, where $k$ is treated as a constant.

\smallskip\


\end{abstract}

\section{Introduction}

Entanglement is a fundamental quantum mechanical resource that can
be shared among spatially separated parties. The possibility of
having entanglement is a distinguishing feature of quantum
mechanics that does not exist in classical mechanics. It plays a
central role in some striking applications of quantum computation
and quantum information such as quantum
teleportation~\cite{BBC+93}, quantum superdense coding~\cite{BS92}
and quantum cryptography~\cite{BB84}. As a result, entanglement
has been recognized as a useful physical resource~\cite{M00}.
However, many fundamental problems concerning quantum entanglement
are still unsolved. An important such problem concerns the
existence of entanglement transformation. Suppose that Alice and
Bob each have one part of a bi-partite  state. The question then
is what other states can they transform the entangled state into?
Since an entangled state is separated spatially, it is natural to
require that Alice and Bob can only make use of local operations
and classical communication (LOCC). Significant progress in the
study of entanglement was made by Bennett, Bernstein, Popescu and
Schumacher~\cite{BBPS96} in 1996. They proposed an entanglement
concentration protocol which solved the entanglement
transformation problem in the asymptotic case. In 1999,
Nielsen~\cite{Nielsen} made another important advance. Suppose
there is a bi-partite state
$|\psi_1\rangle=\sum_{i=1}^{n}\sqrt{\alpha_i}|i\rangle_A|i\rangle_B$
shared between Alice and Bob, with ordered Schmidt coefficients
(OSCs for short) $\alpha_1\geq \alpha_2\geq \cdots \geq
\alpha_n\geq0$, and they want to transform $|\psi_1\rangle$ into
another bi-partite state
$|\psi_2\rangle=\sum_{i=1}^{n}\sqrt{\beta_i}|i\rangle_A|i\rangle_B$
with  OSCs $\beta_1\geq \beta_2\geq \cdots \geq \beta_n\geq0$. It
was proved that $|\psi_1\rangle\rightarrow|\psi_2\rangle$ is
possible under LOCC if and only if $\lambda_{\psi_1}\prec
\lambda_{\psi_2}$, where $\lambda_{\psi_1}$ and $\lambda_{\psi_2}$
are the vectors of ordered Schmidt coefficients, i.e.
$\lambda_{\psi_1}=(\alpha_1,\ldots,\alpha_n)$,
$\lambda_{\psi_2}=(\beta_1,\ldots,\beta_n)$,
$\prec$ denotes the majorization relation~\cite{MO79, AU82}, i.e.
for $1\leq l\leq n,$
$$\sum_{i=1}^{l}\alpha_i\leq \sum_{i=1}^{l}\beta_i,$$ with
equality when $l=n$. This fundamental contribution by Nielsen
provides us with an extremely useful mathematical tool for
studying entanglement transformation. A simple but significant
fact implied by Nielsen's theorem is that there exist incomparable
states $|\psi_1\rangle$ and $|\psi_2\rangle$ with both
transformations $|\psi_1\rangle\rightarrow|\psi_2\rangle$ and
$|\psi_2\rangle\rightarrow|\psi_1\rangle$ impossible. Shortly
after Nielsen's work, a quite surprising phenomenon of
entanglement, namely, entanglement catalysis, was discovered by
Jonathan and Plenio~\cite{Jonathan}. They gave an example showing
that one may use another entangled state $|c\rangle$, known as a
catalyst, to make an impossible transformation $|\psi\rangle
\rightarrow |\phi\rangle$ possible. Furthermore, the
transformation is in fact one of $|\psi\rangle\otimes|c\rangle
\rightarrow |\phi\rangle\otimes|c\rangle$, so that the catalyst
$|c\rangle$ is not modified in the process.

Entanglement catalysis is another useful protocol that quantum
mechanics provides. Therefore to exploit the full power of quantum
information processing, we first have to solve the following basic
problem: given a pair of incomparable states $|\psi_1\rangle$ and
$|\psi_2\rangle$ with
$|\psi_1\rangle\not\rightarrow|\psi_2\rangle$ and
$|\psi_2\rangle\not\rightarrow|\psi_1\rangle$, determine whether
there exists a catalyst $|c\rangle$ such that
$|\psi_1\rangle\otimes|c\rangle \rightarrow
|\psi_2\rangle\otimes|c\rangle$. According to Nielsen's theorem,
solving the problem requires determining whether there is a state
$|c\rangle$ for which the majorization relation
$\lambda_{\psi_1\otimes c}\prec \lambda_{\psi_2\otimes c}$ holds.
As pointed out by Jonathan and Plenio~\cite{Jonathan}, it is very
difficult to find an analytical and both  necessary and sufficient
condition for the existence of a catalyst. The difficulty is
mainly due to lack of suitable mathematical tools to deal with
majorization of tensor product states, and especially the flexible
ordering of the OSCs of tensor products. In ~\cite{Jonathan},
Jonathan and Plenio only gave some simple necessary conditions for
the existence of catalysts, but no sufficient condition was found.
Those necessary conditions enabled them to show that entanglement
catalysis can happen in the transformation between two $n\times n$
states with $n\geq 4$. One of the main aims of the present paper
is to give a necessary and sufficient condition for entanglement
catalysis in the simplest case of entanglement transformation
between $4\times 4$ states with a $2\times 2$ catalyst. For
general case, the fact that an analytical condition under which
incomparable states are catalyzable is not easy to find leads us
naturally to an alternative approach; that is, to seek some
efficient algorithm to decide catalyzability of entanglement
transformation. Indeed, an algorithm to decide the existence of
catalysts was already presented by Bandyopadhyay and
Roychowdhury~\cite{Bandyopadhyay}. Unfortunately, for two $n\times
n$ incomparable states, to determine whether there exists a
$k\times k$ catalyst for them, their algorithm runs in exponential
time with complexity $O([(nk)!]^2)$, and so it is intractable in
practice. The intractability of Bandyopadhyay and Roychowdhury's
algorithm stimulated us to find a more efficient algorithm for the
same purpose, and this is exactly the second aim of the present
paper.

This paper is organized as follows. In the second section we deal
with entanglement catalysis in the simplest case of $n=4$ and $
k=2$. A necessary and sufficient condition under which a $2\times
2$ catalyst exists for an entanglement transformation between
$4\times 4$ states is presented. This condition is analytically
expressed in terms of the OSCs of the states involved in the
transformation, and thus it is easily checkable. Also, some
interesting examples are given to illustrate the use of this
condition. The third section considers the general case. We
propose a polynomial time algorithm to decide the existence of
catalysts. Suppose $|\psi_1\rangle$ and $|\psi_2\rangle$ are two
given $n\times n$ incomparable states, and $k$ is any fixed
natural number. With the aid of our algorithm, one can quickly
find all $k\times k$ catalysts for the transformation
$|\psi_1\rangle\rightarrow|\psi_2\rangle$ using only
$O(n^{2k+3.5})$ time. Comparing to the time complexity
$O([(nk)!]^2)$ of the algorithm given in~\cite{Bandyopadhyay}, for
constant $k$, our algorithm improves the complexity from
superexponential to polynomial. We make conclusions in section 4,
and some open problem are also discussed.

To simplify the presentation, in the rest of the paper, we
identify the state
$|\psi\rangle=\sum_{i=1}^{n}\sqrt{\gamma_i}|i\rangle|i\rangle$
with the vector of its Schmidt coefficients
$(\gamma_1,\gamma_2,\ldots,\gamma_n)$, the meaning will be clear
from the context.

\section{A necessary and sufficient condition of entanglement catalysis in the simplest case ($n=4,k=2$)}

Jonathan and Plenio [9] has shown that entanglement catalysis only
occurs in transformations between $n\times n$ states with $n\geq
4$. In this section, we consider the simplest case that a
transformation from one $4\times 4$ state to another possesses a
$2\times 2$ catalyst. Assume
$|\psi_1\rangle=(\alpha_1,\alpha_2,\alpha_3,\alpha_4)$ and
$|\psi_2\rangle=(\beta_1,\beta_2,\beta_3,\beta_4)$ are two
$4\times 4$ states, where $\alpha_1\geq \alpha_2\geq \alpha_3 \geq
\alpha_4\geq 0$, $\sum_{i=1}^{4}\alpha_i=1$, $\beta_1\geq
\beta_2\geq \beta_3 \geq \beta_4\geq 0$, and
$\sum_{i=1}^{4}\beta_i=1$. The potential catalyst is supposed to
be a $2\times 2$ state, denoted by $|\phi\rangle=(c,1-c)$, where
$c\in [0.5,1]$.

It was proved in~\cite{Jonathan} that if
$|\psi_1\rangle\not\rightarrow |\psi_2\rangle$, but
$|\psi_1\rangle\otimes|\phi\rangle\rightarrow
|\psi_2\rangle\otimes|\phi\rangle$ then
\begin{equation}\label{eq21}
\alpha_1\leq \beta_1, \ \ \alpha_1+\alpha_2>\beta_1+\beta_2,\ \
\alpha_1+\alpha_2+\alpha_3\leq \beta_1+\beta_2+\beta_3,
\end{equation}
 or
equivalently,
\begin{equation}\label{eq22}
\alpha_2+\alpha_3+\alpha_4\geq \beta_2+\beta_3+\beta_4,\ \ \alpha_3+\alpha_4<\beta_3+\beta_4,
\ \ \alpha_4\geq \beta_4.
\end{equation}
Note that $\{\alpha_i\}$ and $\{\beta_i\}$ are arranged in
decreasing order, so we have
\begin{equation}\label{eq23}
\beta_1\geq \alpha_1\geq \alpha_2>\beta_2\geq \beta_3>\alpha_3\geq
\alpha_4\geq \beta_4
\end{equation}
These inequalities are merely necessary conditions for the
existence of catalyst $|\phi\rangle$, and it is easy to see that
they are not sufficient. In the following theorem we give a
condition which is both necessary and sufficient.

\begin{theo}\label{theo21}
There exists a catalysts $|\phi\rangle$ for two states
$(|\psi_1\rangle,|\psi_2\rangle)$ with
$|\psi_1\rangle\not\rightarrow |\psi_2\rangle$, if and only if
\begin{equation}\label{eq24}
\max\left\{\frac{\alpha_1+\alpha_2-\beta_1}{\beta_2+\beta_3},1-\frac{\alpha_4-\beta_4}{\beta_3-\alpha_3}\right\}
\leq \min\left\{\frac{\beta_1}{\alpha_1+\alpha_2},
\frac{\beta_1-\alpha_1}{\alpha_2-\beta_2},1-\frac{\beta_4}{\alpha_3+\alpha_4}\right\}
\end{equation}
and Eq.~(\ref{eq21}) hold. In addition, for any $c\in [0.5,1]$
such that
\begin{equation*}\label{eq25}
\max\left\{\frac{\alpha_1+\alpha_2-\beta_1}{\beta_2+\beta_3},1-\frac{\alpha_4-\beta_4}{\beta_3-\alpha_3}\right\}
\leq c\leq \min\left\{\frac{\beta_1}{\alpha_1+\alpha_2},
\frac{\beta_1-\alpha_1}{\alpha_2-\beta_2},1-\frac{\beta_4}{\alpha_3+\alpha_4}\right\}
\end{equation*}
$|\phi\rangle=(c,1-c)$ is a catalyst for
$(|\psi_1\rangle,|\psi_2\rangle)$.
\end{theo}

\noindent \textit{Proof:} Assume $|\psi_1\rangle\not\rightarrow
|\psi_2\rangle$ but $|\psi_1\rangle\otimes|\phi\rangle\rightarrow
|\psi_2\rangle\otimes|\phi\rangle$ under LOCC. From Eq. (8)
in~\cite{Jonathan} we know Eq.~(\ref{eq21}) holds. So
Eq.~(\ref{eq22}) and Eq.~(\ref{eq23}) hold too.

A routine calculation shows that the Schmidt coefficients of
$|\psi_1\rangle|\phi\rangle$ and $|\psi_2\rangle|\phi\rangle$ are
$$A=\{\alpha_1 c,\alpha_2 c,\alpha_3 c,\alpha_4 c;\alpha_1
(1-c),\alpha_2 (1-c),\alpha_3 (1-c),\alpha_4 (1-c)\}$$ and
$$B=\{\beta_1 c,\beta_2 c,\beta_3 c,\beta_4 c;\beta_1 (1-c),\beta_2
(1-c),\beta_3 (1-c),\beta_4 (1-c)\},$$ respectively. Sort the
elements in $A$ and $B$ in decreasing order and denote the
resulted sequences by $a^{(1)}\geq a^{(2)}\geq \cdots\geq a^{(8)}$
and $b^{(1)}\geq b^{(2)}\geq \cdots\geq b^{(8)}$. It is clear that
$a^{(1)}=\alpha_1 c$, $a^{(8)}=\alpha_4 (1-c)$, $b^{(1)}=\beta_1
c$, and $b^{(8)}=\beta_4 (1-c)$. Since
$|\psi_1\rangle\otimes|\phi\rangle\rightarrow
|\psi_2\rangle\otimes|\phi\rangle$, Nielsen's theorem tells us
that
$$\sum_{i=1}^{l}a^{(i)}\leq \sum_{i=1}^{l}b^{(i)}\ \ (\forall 1\leq
l\leq 8)$$

Since $\{\beta_i\}$ is ordered, and $c\geq 0.5$, thus
\begin{equation}\label{EQbeta1}
\beta_1 c\geq \beta_2 c\geq \beta_3 c\geq \beta_4 c, \ \ \beta_1
(1-c)\geq \beta_2 (1-c) \geq \beta_3 (1-c)  \geq \beta_4 (1-c),\ \
\beta_i c\geq \beta_i (1-c)
\end{equation}
Now we are going to demonstrate that
\begin{equation}\label{EQbeta}
\beta_1 c\geq \beta_1 (1-c) > \beta_2 c\geq \beta_3 c>\beta_2
(1-c) \geq \beta_3 (1-c) > \beta_4 c \geq \beta_4 (1-c).
\end{equation}
and consequently fix the ordering of $B$. The key idea is: the sum
of the biggest $l$ numbers in a set is greater than or equal the
sum of any $l$ numbers in this set.

First, by definition of $\{a^{(i)}\}$ we have $a^{(1)}+a^{(2)}\geq
\alpha_1 c+\alpha_2 c$. So Nielsen's theorem leads to
$b^{(1)}+b^{(2)}\geq a^{(1)}+a^{(2)}\geq \alpha_1 c+\alpha_2 c$.
From inequality~(\ref{eq21}), $\alpha_1+\alpha_2>\beta_1+\beta_2$,
so $b^{(1)}+b^{(2)}>\beta_1 c+\beta_2 c$, i.e. $b^{(2)}>\beta_2
c$. Combining this with inequality~(\ref{EQbeta1}), we see that
the only case is $b^{(2)}=\beta_1 (1-c)$, $b^{(3)}=\beta_2 c$ and
$\beta_1 (1-c)>\beta_2 c$.

Similarly, we have $$a^{(1)}+a^{(2)}+a^{(3)}+a^{(4)}\geq \alpha_1
c+\alpha_2 c+\alpha_1 (1-c)+\alpha_2 (1-c)=\alpha_1+\alpha_2.$$ So
it holds that
$$b^{(1)}+b^{(2)}+b^{(3)}+b^{(4)}\geq
a^{(1)}+a^{(2)}+a^{(3)}+a^{(4)}\geq \alpha_1+\alpha_2>
\beta_1+\beta_2.$$ This implies $b^{(4)}>\beta_2 (1-c).$ Then it
must be that $b^{(4)}=\beta_3 c$, and $\beta_3 c> \beta_2 (1-c)$.

Now what remains is to determine the order between $b^{(5)}$ and
$b^{(7)}$. We consider $b^{(7)}$ first. Nielsen's theorem yields
$b^{(7)}+b^{(8)}\leq a^{(7)}+a^{(8)}$. By definition, we know that
$a^{(7)}+a^{(8)}\leq \alpha_3 (1-c) + \alpha_4 (1-c)$. Therefore,
$$b^{(7)}+b^{(8)}\leq \alpha_3 (1-c) + \alpha_4
(1-c)=(\alpha_3+\alpha_4)(1-c)<(\beta_3+\beta_4)(1-c),$$ the last
inequality is due to~(\ref{eq22}). Since $b^{(8)}=\beta_4 (1-c)$,
it follows that $b^{(7)}<\beta_3(1-c)$. Furthermore, we obtain
$b^{(7)}=\beta_4 c, b^{(6)}=\beta_3(1-c)$, and
$\beta_3(1-c)>\beta_4 c$.

Finally, only $\beta_2 (1-c)$ leaves, so $b^{(5)}=\beta_2 (1-c)$.
Combining the above arguments, we finish the proof of
inequality~(\ref{EQbeta}).

Clearly, inequality~(\ref{EQbeta}) implies that
\begin{equation}\label{eq0}
\frac{\beta_2}{\beta_2+\beta_3}<c<\left\{\frac{\beta_1}{\beta_1+\beta_2},\frac{\beta_3}{\beta_3+\beta_4}\right\}
\end{equation}
This is needed in the remainder of the proof.

Remembering the order of $B$ has been found out, it enables us to
calculate easily $\sum_{i=1}^{l}b^{(i)}$ for each $l$. The only
rest problem is how to calculate $\sum_{i=1}^{l}a^{(i)}$. To this
end, we need the following simple lemma:

\begin{lemm}\label{lemm22}
Assume $A=\{a_1,\ldots,a_n\}$, $B=\{b_1,\ldots,b_n\}$. Sort $B$ in
decreasing order and denote the resulted sequence by $b^{(1)}\geq
b^{(2)}\geq \cdots\geq b^{(n)}$. Then $A\prec B$ if and only if
for $1\leq l \leq n$,
\begin{equation}\label{eq2}
\max_{A'\subseteq A, |A'|=l}\sum_{a_i\in A'}a_i\leq
\sum_{i=1}^{l}b^{(i)}
\end{equation}
with equality when $l=n$.
\end{lemm}

\noindent \textit{Proof of Lemma:} The \lq\lq if" part is obvious.
For the \lq\lq only if " part, we sort $A$ in decreasing order and
denote the resulted sequence by $a^{(1)}\geq a^{(2)}\geq
\cdots\geq a^{(n)}$. Then $A\prec B$ if and only if for $1\leq l
\leq n$,
$$\sum_{i=1}^{l}a^{(i)}\leq \sum_{i=1}^{l}b^{(i)}$$
It is easy to see that $\displaystyle
\sum_{i=1}^{l}a^{(i)}=\max_{A'\subseteq A, |A'|=l}\sum_{a_i\in
A'}a_i$, so the lemma holds.

\noindent \textit{Proof of Theorem 2.1 (continued):} Now the above
lemma guarantees a quite easy way to deal with
$\sum_{i=1}^{l}a^{(i)}$: enumerating simply all the possible
cases. For example, $a^{(1)}+a^{(2)}=\alpha_1 c+\alpha_1 (1-c)$ or
$\alpha_1 c+\alpha_2 c$, i.e. $a^{(1)}+a^{(2)}=\max\{\alpha_1
c+\alpha_1 (1-c),\alpha_1 c+\alpha_2 c\}$. The treatments for
$\sum_{i=1}^{3}a^{(i)},\ldots,\sum_{i=1}^{8}a^{(i)}$ are the same.
What we still need to do now is to solve systematically the
inequalities of $\sum_{i=1}^{l}a^{(i)}\leq \sum_{i=1}^{l}b^{(i)}$
$(1\leq l\leq 8)$. We put this daunting but routine part in the
Appendix.\hfill$\Box$

\vspace{0.2cm}

The above theorem presents a necessary and sufficient condition
when a $2\times 2$ catalyst exists for a transformation from one
$4\times 4$ state to another. Moreover, it is also worth noting
that the theorem is indeed constructive. The second part of it
gives all $2\times 2$ catalysts (if any) for such a
transformation. To illustrate the utility of the above theorem,
let us see some simple examples.


\begin{exam} \upshape This example is exactly the original example
that Jonathan and Plenio~\cite{Jonathan} used to demonstrate
entanglement catalysis. Let $|\psi_1\rangle=(0.4,0.4,0.1,0.1)$ and
$|\psi_2\rangle=(0.5,0.25,0.25,0)$. Then
$$\max\left\{\frac{\alpha_1+\alpha_2-\beta_1}{\beta_2+\beta_3},1-\frac{\alpha_4-\beta_4}{\beta_3-\alpha_3}\right\}
=\max\{0.6,1-2/3\}=0.6,$$
$$\min\left\{\frac{\beta_1}{\alpha_1+\alpha_2},
\frac{\beta_1-\alpha_1}{\alpha_2-\beta_2},1-\frac{\beta_4}{\alpha_3+\alpha_4}\right\}=\min\{5/8,2/3,1-0\}=0.625.$$
Since $0.6<0.625$, Theorem 2.1 gives us a continuous spectrum
$|\phi\rangle=(c,1-c)$ of catalysts for $|\psi_1\rangle$ and
$|\psi_2\rangle$, where $c$ ranges over the interval
$[0.6,0.625]$. Especially, when choosing $c=0.6$, we get the
catalyst $|\phi\rangle=(0.6,0.4)$, which is the one given
in~\cite{Jonathan}.
\end{exam}


\begin{exam}\upshape We also consider the example
in~\cite{Bandyopadhyay}. Let $|\psi_1\rangle=(0.4,0.36,0.14,0.1)$
and $|\psi_2\rangle=(0.5,0.25,0.25,0)$. The catalyst for
$|\psi_1\rangle$ and $|\psi_2\rangle$ given there is
$\phi=(0.65,0.35)$. Note that
$$\max\left\{\frac{\alpha_1+\alpha_2-\beta_1}{\beta_2+\beta_3},1-\frac{\alpha_4-\beta_4}{\beta_3-\alpha_3}\right\}
=\max\{0.52,1-10/11\}=0.52,$$
$$\min\left\{\frac{\beta_1}{\alpha_1+\alpha_2},
\frac{\beta_1-\alpha_1}{\alpha_2-\beta_2},1-\frac{\beta_4}{\alpha_3+\alpha_4}\right\}=\min\{25/38,10/11,1-0\}=25/38,$$
and $0.52<0.65<25/38$, Theorem 2.1 guarantees that $|\phi\rangle$
is really a catalyst; and it allows us to find much more catalysts
$|\phi\rangle =(c,1-c)$ with $c\in [0.52,25/38]$.
\end{exam}

\section{An efficient algorithm for deciding existence of catalysts}

In the last section, we was able to give a necessary and
sufficient condition under which a $2\times 2$ catalyst exists for
an transformation between $4\times 4$ states. The key idea
enabling us to obtain such a condition is that the order among the
Schmidt  coefficients of the tensor product of the catalyst and
the target state in the transformation is uniquely determined by
Nielsen's Theorem.
However, the same idea does not work when we deal with higher
dimensional states, and it seems very hard to find an analytical
condition for existence of catalyst in the case of higher
dimension. On the other hand, existence of catalysts is a dominant
problem in exploiting the power of entanglement catalysis in
quantum information processing. Such a dilemma forces us to
explore alternatively the possibility of finding an efficient
algorithm for deciding existence of catalysts. The main purpose is
to give a polynomial time algorithm to decide whether there is a
$k\times k$ catalyst for two incomparable $n\times n$ states
$|\psi_1\rangle,|\psi_2\rangle$, where $k\geq2$ is a fixed natural
number.

To explain the intuition behind our algorithm more clearly, we
first cope with the case of $k=2$. Assume
$|\psi_1\rangle=(\alpha_1,\ldots,\alpha_n)$, and
$|\psi_2\rangle=(\beta_1,\ldots,\beta_n)$ are two $n\times n$
states, and assume that the potential catalyst for them is a
$2\times 2$ state $\phi=(x,1-x)$. The Schmidt coefficients of
$|\psi_1\rangle|\phi\rangle$ and $|\psi_2\rangle|\phi\rangle$ are
then given as $$A_x=\{\alpha_1 x,\alpha_2 x,\ldots,\alpha_n x;
\alpha_1 (1-x),\ldots,\alpha_n (1-x)\}$$ and $$B_x=\{\beta_1
x,\beta_2 x,\ldots,\beta_n x; \beta_1 (1-x),\ldots,\beta_n
(1-x)\},$$ respectively. Sort them in decreasing order and denote
the resulting sequences by $a^{(1)}(x)\geq a^{(2)}(x)\geq
\cdots\geq a^{(2n)}(x)$ and $b^{(1)}(x)\geq b^{(2)}(x)\geq
\cdots\geq b^{(2n)}(x)$. By Nielsen's theorem we know that a
necessary and sufficient condition for
$|\psi_1\rangle|\phi\rangle\rightarrow|\psi_2\rangle|\phi\rangle$
is $$\sum_{i=1}^{l}a^{(i)}(x)\leq \sum_{i=1}^{l}b^{(i)}(x)\
(l=1,\ldots,2n).$$ Now the difficulty arises from the fact that we
do not know the exact order of elements in $A$ and $B$. Let us now
consider this problem in a different way. If we fix $x$ to some
constant $x_0$, we can calculate the elements in $A$, $B$ and sort
them. Then if we moves $x$ slightly from $x_{0}$ to
$x_0+\epsilon$, the order of the elements in $A$ (or $B$) does not
change, except the case that $x$ goes through a point $x^{*}$ with
$\alpha_i(1-x^{*})=\alpha_j x^{*}$ (or $\beta_i (1-x^{*})=\beta_j
x^{*}$), i.e. $x^{*}=\frac{\alpha_i}{\alpha_i+\alpha_j}$ (resp.
$x^{*}=\frac{\beta_i}{\beta_i+\beta_j}$) for some $i<j$. This
observation leads us to the following algorithm:

\begin{center}
\fbox{
\begin{minipage}[ht]{15cm}

\textbf{Algorithm 1}

%

\begin{enumerate}
\item
$\rho_{i,j}\leftarrow\frac{\alpha_i}{\alpha_i+\alpha_j},\delta_{i,j}\leftarrow\frac{\beta_i}{\beta_i+\beta_j},\
1\leq i<j\leq n$

\item Sort $\{\rho_{i,j}\}\cup\{\delta_{i,j}\}$ in nondecreasing
order, the resulted sequence is denoted by $\gamma^{(1)}\leq
\gamma^{(2)} \leq \cdots\leq \gamma^{(n^2-n)}$

\item $\gamma^{(0)}\leftarrow 0.5,\gamma^{(n^2-n+1)}\leftarrow 1$

\item \textbf{For} $i=0$ \textbf{to} $n^2-n$ \textbf{do}

\item \hspace{0.6cm}
$c\leftarrow\frac{\gamma^{(i)}+\gamma^{(i+1)}}{2}$

\item \hspace{0.6cm} Determine the order of elements in $A_c$ and
$B_c$, respectively

\item \hspace{0.6cm} Solve the system of inequalities:
$$\left\{
\begin{array}{ll}
\sum_{i=1}^{l}a^{(i)}(x)\leq \sum_{i=1}^{l}b^{(i)}(x) &
(l=1,\ldots,2n)\\ \gamma^{(i)}\leq x\leq \gamma^{(i+1)} &
\end{array}
\right.$$

\item \textbf{OUTPUT}: Catalysts do not exist, if for all $i\in
\{0,1,\ldots,n^2-n\}$, the solution set of the above inequalities
is empty; catalyst $(x,1-x)$, if for some $i$ the inequalities has
solution.
\end{enumerate}

\end{minipage}
}
\end{center}

It is easy to see that this algorithm runs in $O(n^3)$ time. In
\cite{Bandyopadhyay}, an algorithm for the same purpose was also
given, but it runs in $O(n!)$ time.

By generalizing the idea explained above to the case of $k\times
k$ catalyst, we obtain:

\begin{theo}
For any two $n\times n$ states
$|\psi_1\rangle=(\alpha_1,\ldots,\alpha_n)$ and
$|\psi_2\rangle=(\beta_1,\ldots,\beta_n)$, the problem whether
there exists a $k\times k$ catalyst
$|\phi\rangle=(x_1,\ldots,x_k)$ for them can be decided in
polynomial time about $n$. Further more, if there exists a
$k\times k$ catalyst, our algorithm can find all the catalysts in
$O(n^{2k+3.5})$ time.
\end{theo}

\noindent \textit{Proof.} The algorithm is similar to the one for
the case $k=2$. Now the Schmidt coefficients of
$|\psi_1\rangle|\phi\rangle$ and $|\psi_2\rangle|\phi\rangle$ are
$$A_x=\{\alpha_1 x_1,\ldots,\alpha_n x_1; \alpha_1
x_2,\ldots,\alpha_n x_2;\ldots,\alpha_n x_k\}$$ and
$$B_x=\{\beta_1 x_1,\ldots,\beta_n x_1; \beta_1 x_2,\ldots,\beta_n
x_2;\ldots,\beta_n x_k\}.$$ If we move $x$ in the $k-$dimensional
space $\mathbb{R}^k$, the order of the elements in $A_x$ (or
$B_x$) will change if and only if $x$ goes through a hyperplane
$\alpha_{i_1} x_{i_2}=\alpha_{j_1}x_{j_2}$ ($\beta_{i_1}
x_{i_2}=\beta_{j_1}x_{j_2}$) for some $i_1<j_1$ and $i_2>j_2$.
(Indeed, the area that $x$ ranges over should be
$(k-1)-$dimensional because we have a constrain of
$\sum_{i=1}^{k}x_i=1$.) So first we can write down all the
equations of these hyperplanes $$\Gamma=\{\alpha_{i_1}
x_{i_2}=\alpha_{j_1}x_{j_2}|i_1<j_1,i_2>j_2\}\cup \{\beta_{i_1}
x_{i_2}=\beta_{j_1}x_{j_2}|i_1<j_1,i_2>j_2\},$$ where
$|\Gamma|=2{k \choose 2}{n \choose 2}=O(n^2)$. In the
$k-$dimensional space $\mathbb{R}^{k}$, these $O(n^2)$ hyperplanes
can at most divide the whole space into $O(O(n^2)^k)=O(n^{2k})$
different parts. Note the number of parts generated by these
hyperplanes is a polynomial of $n$. Now we enumerate all these
possible parts. In each part, for different $x$, the elements in
$A_x$ (or $B_x$) has the same order. Then we can solve the
inequalities $$\sum_{i=1}^{l}a^{(i)}(x)\leq
\sum_{i=1}^{l}b^{(i)}(x)\ (1\leq l \leq nk)$$ and check the order
constrains by linear programming. Following the well-known result
that linear programming is solvable in $O(n^{3.5})$ time, our
algorithm runs in $O(n^{2k+3.5})$ time, it is a polynomial time of
$n$ whenever $k$ is a given constant.\hfill $\Box$

Indeed, Theorem 3.1 is constructive too, and its proof gives an
algorithm which is able not only to decide whether a catalyst of a
given dimension exists but also to find all such catalysts when
they do exist. The algorithm before this theorem is just a more
explicit presentation of the proof for the case of $k=2$.

\section{Conclusion and discussion}

In this paper, we investigate the problem concerning existence of
catalysts for entanglement transformations. It is solved for the
simplest case in an analytical way. We give a necessary and
sufficient condition for the existence of a $2\times 2$ catalyst
for a pair of two incomparable $4\times 4$ states. For the general
case ($k\times k$ catalysts for $n\times n$ states), although we
fail to give an analytical condition, an efficient polynomial time
algorithm is found when $k$ is treated as a constant. However, if
$k$ is a variable, ranging over all positive integers, the problem
of determining the existence of catalysts still remains open. We
believe it is NP-hard, since the set $A_x=\{\alpha_1
x_1,\ldots,\alpha_n x_1; \alpha_1 x_2,\ldots,\alpha_n
x_2;\ldots,\alpha_n x_k\}$ in the proof of Theorem 3.1 potentially
has exponential kind of different orders.

\smallskip\

\textbf{Acknowledgements}: The authors are very grateful to the
anonymous referees for their invaluable comments and suggestions
that  helped to improve the presentation in this paper.

\section{Appendix: Proof of Theorem~\ref{theo21}}
\noindent \textit{Proof of Theorem 2.1 (remaining part)}: We need
to solve the system of inequalities $\sum_{i=1}^{l}a^{(i)}\leq
\sum_{i=1}^{l}b^{(i)}$ $(1\leq l\leq 8)$. This is carried out by
the following items:

(1) First, we have:

\begin{equation}\label{eq1}
 a^{(1)}\leq b^{(1)} \ \ \Longleftrightarrow \ \  \alpha_1 c \leq
\beta_1 c \ \ \Longleftrightarrow \ \ \alpha_1 \leq \beta_1.
\end{equation}

(2) The inequality $a^{(1)}+a^{(2)}\leq  b^{(1)}+b^{(2)}$ may be
rewritten as
\begin{eqnarray}
\max\{\alpha_1 c+\alpha_1 (1-c),\alpha_1 c+\alpha_2 c\} \leq
\beta_1 c+\beta_1 (1-c) & \Longleftrightarrow & \\
 c & \leq & \frac{\beta_1}{\alpha_1+\alpha_2},
\ \ \alpha_1 \leq \beta_1.
\end{eqnarray}

(3) We now consider $a^{(1)}+a^{(2)}+a^{(3)}\leq
b^{(1)}+b^{(2)}+b^{(3)}$. It is equivalent to
\begin{eqnarray}
  \max\{\alpha_1 c+\alpha_1 (1-c)+\alpha_2 c,\alpha_1 c+\alpha_2
c+\alpha_3 c\}& \leq &
\beta_1 c+\beta_1 (1-c)+\beta_2 c \ \ \ \Longleftrightarrow \nonumber \\
  c & \leq &
\left\{\frac{\beta_1}{\alpha_1+\alpha_2+\alpha_3-\beta_2},\frac{\beta_1-\alpha_1}{\alpha_2-\beta_2}\right\}
\end{eqnarray}

(4) It holds that
\begin{eqnarray}
 a^{(1)}+a^{(2)}+a^{(3)}+a^{(4)}\leq
b^{(1)}+b^{(2)}+b^{(3)}+b^{(4)}
 & \Longleftrightarrow &  \nonumber \\
 \max\{\alpha_1 c+\alpha_1 (1-c)+\alpha_2 c+\alpha_2
(1-c),\alpha_1
c+\alpha_2 c+\alpha_3 c+\alpha_1(1-c),& {} & \nonumber \\
 \alpha_1 c+\alpha_2 c+\alpha_3 c+\alpha_4 c\}  \leq
\beta_1 c+\beta_1 (1-c)+\beta_2 c +\beta_3 c \ & \Longleftrightarrow & \nonumber \\
 \frac{\alpha_1+\alpha_2-\beta_1}{\beta_2+\beta_3}\leq c\leq
\left\{\frac{\beta_1}{1-\beta_2-\beta_3},
\frac{\beta_1-\alpha_1}{\alpha_2+\alpha_3-\beta_2-\beta_3}\footnotemark[1]\right\}
& {} &
\end{eqnarray}
\footnotetext[1]{if $\alpha_2+\alpha_3-\beta_2-\beta_3\leq 0$,
this term is useless.}

(5)
\begin{eqnarray}
 a^{(1)}+a^{(2)}+a^{(3)}+a^{(4)}+a^{(5)}\leq
b^{(1)}+b^{(2)}+b^{(3)}+b^{(4)}+b^{(5)}
 & \Longleftrightarrow &   \nonumber \\
 a^{(6)}+a^{(7)}+a^{(8)}\geq b^{(6)}+b^{(7)}+b^{(8)}
 & \Longleftrightarrow &   \nonumber \\
 \min\{\alpha_2 (1-c)+\alpha_3 (1-c)+\alpha_4 (1-c),\alpha_3
(1-c)+\alpha_4 c+\alpha_4(1-c)\} & & \nonumber\\
  \geq
\beta_3 (1-c)+\beta_4 c+\beta_4 (1-c)  & \Longleftrightarrow & \nonumber \\
 1-\frac{\alpha_4-\beta_4}{\beta_3-\alpha_3}\leq c\leq
1-\frac{\beta_4}{\alpha_2+\alpha_3+\alpha_4-\beta_3} & &
\end{eqnarray}

(6)
\begin{eqnarray}
& \sum_{i=1}^{6}a^{(i)}\leq \sum_{i=1}^{6}b^{(i)}
\ & \Longleftrightarrow \   \nonumber \\
& a^{(7)}+a^{(8)}\geq b^{(7)}+b^{(8)}
\ & \Longleftrightarrow \   \nonumber \\
& \min\{\alpha_3 (1-c)+\alpha_4 (1-c),\alpha_4
c+\alpha_4(1-c)\}\geq
\beta_4 c+\beta_4 (1-c) \ & \Longleftrightarrow \ \nonumber \\
&c\leq 1-\frac{\beta_4}{\alpha_3+\alpha_4}, \ \ \alpha_4\geq
\beta_4&
\end{eqnarray}

(7) We have
\begin{equation}\label{eq8}
\sum_{i=1}^{7}a^{(i)}\leq \sum_{i=1}^{7}b^{(i)} \ \
\Longleftrightarrow \ \ a^{(8)}\geq b^{(8)}  \ \
\Longleftrightarrow \ \ \alpha_4\geq \beta_4
\end{equation}

Combining Eq.~(\ref{eq0}, \ref{eq1}-\ref{eq8}) we obtain
\begin{eqnarray}\label{eqC1}
c\leq
\left\{\frac{\beta_1}{\beta_1+\beta_2},\frac{\beta_3}{\beta_3+\beta_4};\frac{\beta_1}{\alpha_1+\alpha_2},\frac{\beta_1}{\alpha_1+\alpha_2+\alpha_3-\beta_2},
\frac{\beta_1-\alpha_1}{\alpha_2-\beta_2},\frac{\beta_1}{1-\beta_2-\beta_3},\right. \nonumber\\
\left.
\frac{\beta_1-\alpha_1}{\alpha_2+\alpha_3-\beta_2-\beta_3}\footnotemark[1],
1-\frac{\beta_4}{\alpha_2+\alpha_3+\alpha_4-\beta_3},1-\frac{\beta_4}{\alpha_3+\alpha_4}\right\}
\end{eqnarray}
and
\begin{equation}\label{eqC2}
c\geq
\left\{\frac{\alpha_1+\alpha_2-\beta_1}{\beta_2+\beta_3},1-\frac{\alpha_4-\beta_4}{\beta_3-\alpha_3}\right\}
\end{equation}
Since
$$\beta_1\geq \alpha_1\geq \alpha_2>\beta_2\geq \beta_3>\alpha_3\geq
\alpha_4\geq \beta_4, \ \ \alpha_1+\alpha_2>\beta_1+\beta_2,
$$
it follows that
$$\frac{\beta_1}{\beta_1+\beta_2}>\frac{\beta_1}{\alpha_1+\alpha_2},$$
$$\frac{\beta_3}{\beta_3+\beta_4}=1-\frac{\beta_4}{\beta_3+\beta_4}>1-\frac{\beta_4}{\alpha_3+\alpha_4},$$
$$\frac{\beta_1}{\alpha_1+\alpha_2}<\frac{\beta_1}{\alpha_1+\alpha_2+(\alpha_3-\beta_2)},$$
$$\frac{\beta_1}{\alpha_1+\alpha_2}<\frac{\beta_1}{\beta_1+\beta_2}<\frac{\beta_1}{\beta_1+\beta_4}
=\frac{\beta_1}{1-\beta_2-\beta_3}$$
$$1-\frac{\beta_4}{\alpha_2+\alpha_3+\alpha_4-\beta_3}>1-\frac{\beta_4}{\alpha_3+\alpha_4},$$
and
$$\frac{\beta_1-\alpha_1}{\alpha_2+\alpha_3-\beta_2-\beta_3}\geq
\frac{\beta_1-\alpha_1}{\alpha_2-\beta_2}.$$ This indicates that
there are six useless terms in Eq.~(\ref{eqC1}), so we can omit
them. Now we get
\begin{equation*}
\max\left\{\frac{\alpha_1+\alpha_2-\beta_1}{\beta_2+\beta_3},1-\frac{\alpha_4-\beta_4}{\beta_3-\alpha_3}\right\}
\leq c\leq \min\left\{\frac{\beta_1}{\alpha_1+\alpha_2},
\frac{\beta_1-\alpha_1}{\alpha_2-\beta_2},1-\frac{\beta_4}{\alpha_3+\alpha_4}\right\}.
\end{equation*}
Therefore, Eq.~(\ref{eq24}) is a necessary condition for the
existence of catalyst.

On the other hand, we claim that Eq.~(\ref{eq21}) and
Eq.~(\ref{eq24}) are the sufficient conditions. Indeed, if we
choose a $c$ satisfies Eq.~(\ref{eq25}), then $c$ satisfies
Eq.~(\ref{eqC1}) and (\ref{eqC2}). From Eq.~(\ref{eq1}-\ref{eq8})
we know that $\sum_{i=1}^{k}a^{(i)}\leq \sum_{i=1}^{k}b^{(i)}$,
i.e. $|\psi_1\rangle|\phi\rangle\rightarrow
|\psi_2\rangle|\phi\rangle$ under LOCC. This completes the proof.
\hfill$\Box$
\end{document}